\documentclass[9pt,twocolumn,twoside]{opticajnl}
\journal{opticajournal} % use for journal or Optica Open submissions

\setboolean{shortarticle}{true}
\usepackage{lineno}
\usepackage{amsmath}
\title{Efficient Light Generation in Ultraviolet-A Band on Chip}

\author[1,2]{Michel Inman}
\author[1,2]{Zhaohui Ma}
\author[1,2]{Zhan Li}
\author[1,2,*]{Yuping Huang}

\affil[1]{Department of Physics, Stevens Institute of Technology, 1 Castle Point Terrace, Hoboken, New Jersey, 07030, USA}
\affil[2]{Center for Quantum Science and Engineering, Stevens Institute of Technology, 1 Castle Point Terrace, Hoboken, New Jersey, 07030, USA}
\affil[*]{Corresponding author: yuping.huang@stevens.edu}

\begin{abstract}
Lithium niobate nanophotonics provides highly efficient nonlinear optics processes covering a broad spectrum from ultraviolet to mid-infrared, yet studies thus far have concentrated in the near-infrared regime. Here we demonstrate light generation in the Ultraviolet-A band in a periodic poled waveguide via second harmonic generation. The internal efficiency reaches 1797 $\% W^{-1}/cm^{-2}$, marking a 9.1-times improvement over the state of art, thanks to better mode overlap and poling. Our technique can find applications in  atomic clocks, frequency comb generation, sensing, and visible entanglement generation.
\end{abstract}

\setboolean{displaycopyright}{false} % Do not include copyright or licensing information in submission.

\begin{document}

\maketitle

\section{Introduction}
Ultraviolet-A (UV-A) band lies to the left edge of the visible spectrum, spanning between 320 nm and 400 nm. It has many applications like; atomic clocks \cite{king2022optical}  entangled photon pair generation \cite{edamatsu2004generation}, , quantum memory \cite{wang2021single}, and frequency comb generation \cite{liu2019beyond}. While UV-A lasers have traditionally been in bulk optics, recent studies have suggested on-chip UV-light generation with advantages in efficiency and integration. To this end, silica waveguides have been used to generate light ranging from 322 nm to 545 nm \cite{yoon2017coherent}, and lithium tantalate microdisks have been used to generate 384 nm light \cite{xue2023chip}, but they suffer low efficiency. \\

Another promising candidate for high efficiency is lithium niobate on insulator (LNOI). Recently, LNOI-based photonic integrated circuits have shown exceptional advantages for optical information processing, due to its wide transparency window and strong Pockels and nonlinear-optical effects \cite{jin2021efficient, lu2019periodically}, and low background noise from two-photon absorption and Raman scattering \cite{fan2021photon}. As such, a variety of utilities have been demonstrated in LNOI, including recent electro-optical modulation \cite{hou2024high,li2020lithium, wang2018nanophotonic}, integrated microwave photonic processing \cite{feng2024integrated}, photon pair generation \cite{shi2024efficient,ma2020ultrabright}, and broadband frequency comb generation \cite{tang2024broadband,wu2024visible}. However, its uses for short-wavelength light generation have been much rarer, with only countable demonstrations at 435 nm \cite{sayem2021efficient}, 456 nm \cite{park2022high}, and most recently in the UV-A band at 355 nm \cite{hwang2023tunable}, although the latter had a low efficiency at 197$\% W^{-1} cm^{-2}$.  

In this paper, we report a 9.1 times improvement in the on chip second harmonic (SH) conversion efficiency in the UV-A band, reaching 1797$\% W^{-1} cm^{-2}$. We use a periodic poled LNOI waveguide that is quasi-phase matched for the second harmonic generation (SHG) of 770.08 nm into 385.04 nm. The higher conversion efficiency is attributed to better poling quality for quasi-phase matching, tighter mode confinement, and nearly ideal mode overlapping. This performance is desirable for expanding LNOI's promises for efficient nonlinear optical processing into the UV-A band and beyond. 

\section{Fabrication}

\begin{figure}[ht]
\centering
\includegraphics[width=0.6\linewidth]{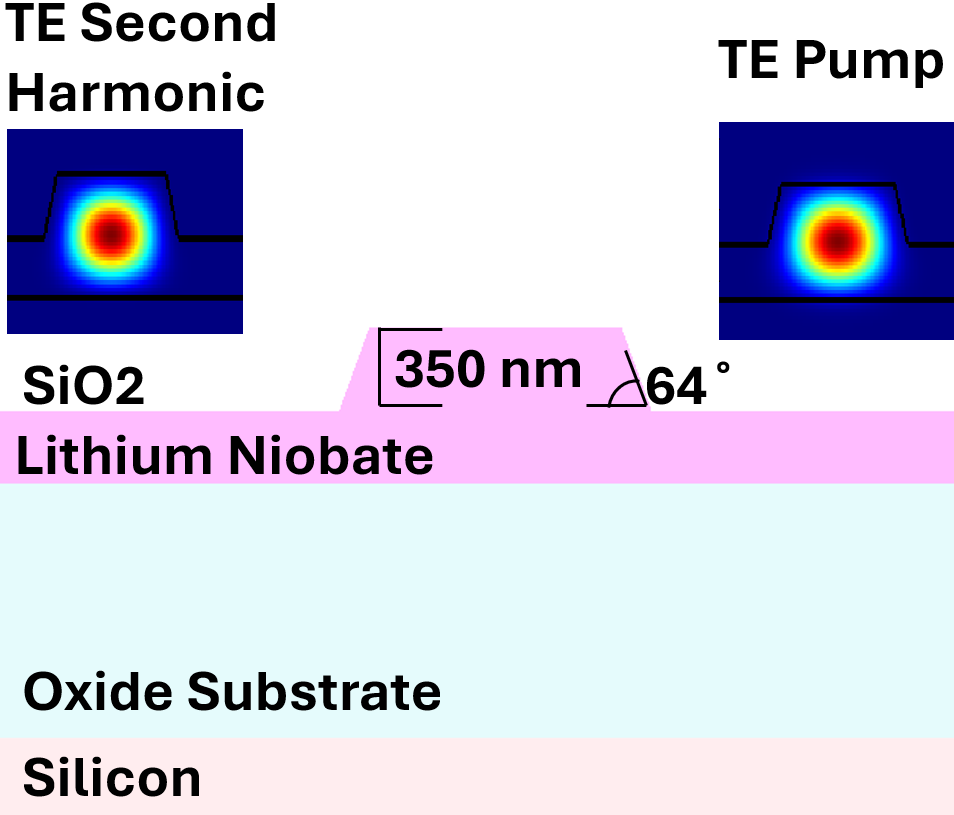}
\caption{Cross-section geometry of the waveguide with the simulated pump and SH mode profiles (both transverse electric).}
\label{fig:Waveguide geometry_modes}
\end{figure}

Figure~\ref{fig:Waveguide geometry_modes} outlines the waveguide's geometry. It is designed with the help of Lumerical [\url{https://www.ansys.com/products/optics/mode}], where the spatial modes are simulated and optimized for second harmonic generation. Using its finite difference eigenmode simulation, we find the feasible top width, sidewall angle, and etching thickness, as well as the poling period. For a top width of 1.4 $\mu$m, a side-wall angle of 64$^\circ$, and an etching depth of 350 nm, the poling period is calculated to be 1.71 $\mu$m for SHG of 780 nm to 390 nm. \\

Next, we use Lumerical's Mode Solver to obtain the transverse electric (TE) modes for both 390 nm and 780 nm waves. As seen in Fig.~\ref{fig:Waveguide geometry_modes}, both modes are well contained in the waveguide. Using these modes, we can compute the conversion efficiency as \cite{chen2019efficient}
\begin{equation} \label{Equation 1}
\eta = \frac{8\pi^2}{\epsilon_0 c \lambda^2_{2\omega}} \, \frac{d_{eff}^2}{n^{2\omega}_{eff} (n^{\omega}_{eff})^2} \, \frac{{\xi}^2}{A_{2\omega}^{2/3} (A_\omega)^{1/3}}\,L^2 \mathrm{sinc}^2\left(\frac{\Delta K L}{2}\right).
\end{equation}
Here, $\epsilon_0$ is the vacuum permittivity. $c$ is the speed of light. $\lambda_{2\omega}$ is the signal wavelength. $L$ is the length of the waveguide (3.5 mm in this case). $d_{eff}$ is $2 d_{33}/\pi$, with $d_{33}$ being the susceptibility of lithium niobate $\chi^{(2)}$. $n^{\omega,2\omega}_{eff}$ is the effective refractive index of each light, retrieved from simulation. $\xi$ is the mode overlap factor, $A_{2\omega}$ and $A_\omega$ are the cross-sectional areas of the mode for SH light and pump light, respectively. They are defined as
\begin{equation} 
\xi=\frac{\iint\,E_{2\omega}^{*}\,{E_{\omega}^2\,dx\,dz}}{{\iint\,|E_{2\omega}|^2E_{2\omega}\,dx\,dz}^{2/3} \,\iint\,{||E_{\omega}|^2}E_{\omega}\,dx\,dz|^{1/3}},
\end{equation}
with the cross-sectional mode areas
\begin{equation}
A_{2\omega}=\frac{\iint\,(|E_{2\omega}|^2\,dx\,dz)^3}{\iint\,(|E_{2\omega}|^2E_{2\omega}\,dx\,dz)^2}
\end{equation}
\begin{equation}
A_\omega=\frac{\iint\,(|E_{\omega}|^2\,dx\,dz)^3}{\iint\,(|E_{\omega}|^2E_{\omega}\,dx\,dz)^2}.
\end{equation}
Here, $E_{\omega, 2\omega}$ are the electric fields of each wave. \\

\begin{figure}[ht]
\centering
\includegraphics[width=\linewidth]{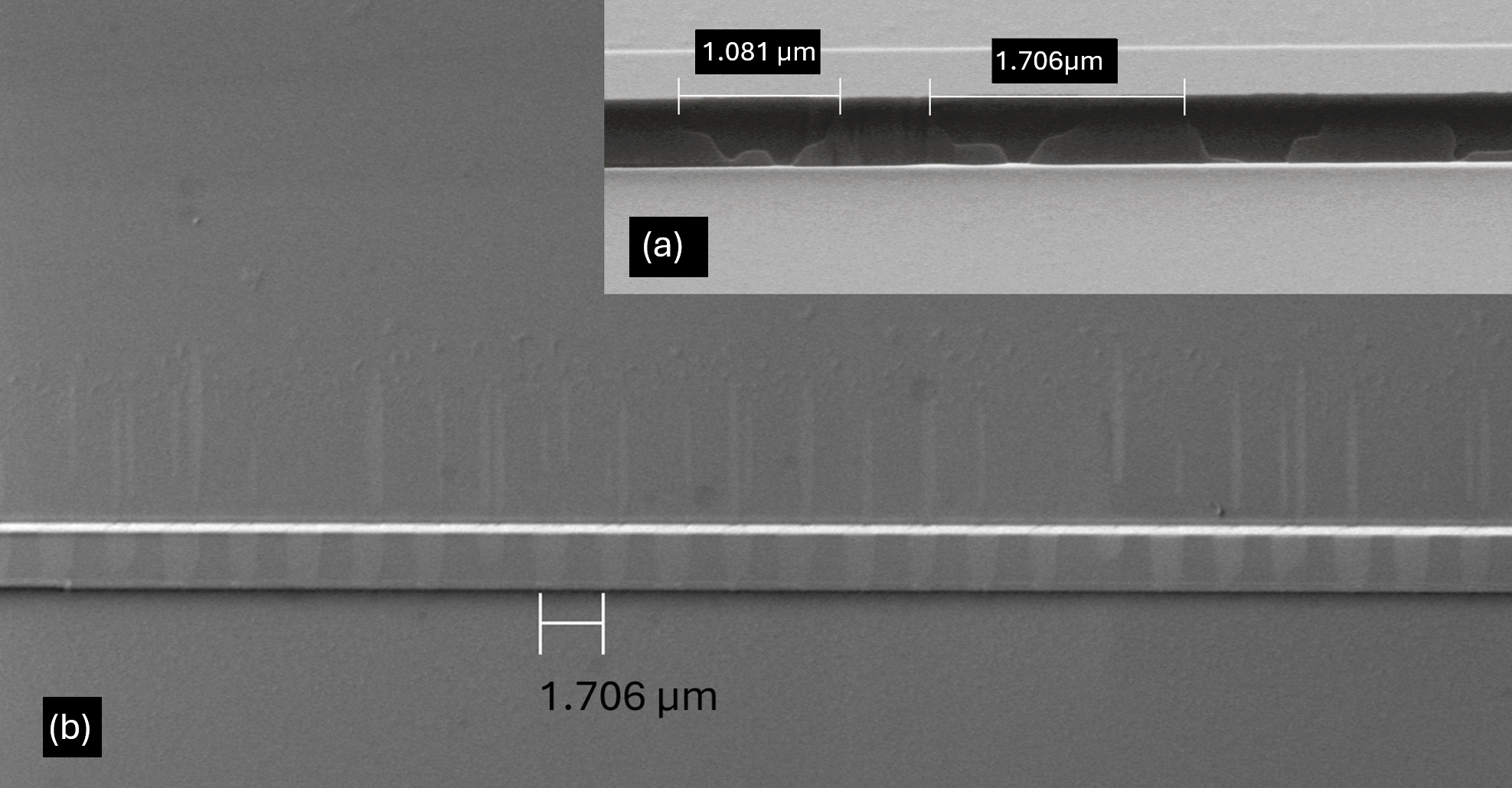}
\caption{An SEM image of the poled LNOI. (a) is the sidewall view to show the duty cycle and poling period. (b) is the top down view showing poled sectors in contrast.}
\label{fig:PPLN Blue Light pole check}
\end{figure}

With the modes in Fig.~\ref{fig:Waveguide geometry_modes}, our calculation finds $A_{2\omega}=0.55\mu m^2$, $A_\omega=0.66\mu m^2$, to give nearly-perfect mode overlap $\xi=0.97$; for details, see \cite{luo2019semi,chen2019efficient}. With the parameters set as above, the theoretical SHG efficiency $\eta$ is 10467 $\% W^{-1} cm^{-2}$. This result is for an ideal poling with a 50:50 duty cycle, where the poled and unpoled sectors are of equal length. However, as seen in Fig.~\ref{fig:PPLN Blue Light pole check}, the poling result is not ideal, showing a 35:65 duty cycle. This leads to a $3.6$-times decrease in the simulated efficiency, to 2905 $\% W^{-1} cm^{-2}$. \\

The waveguide is fabricated on an X-Cut thin film lithium niobate (TFLN) wafer (NANOLN Inc.), which is a 600 nm thick lithium niobate (LN) thin film bonded to a 2 $\mu$m thermally grown silicon dioxide layer above a silicon substrate. The fabrication starts with the electrodes fabricated with a multilayer resist (PMMA 495 A6, 950 A4) spun onto the wafer and its pattern written using 50 keV electron beam lithography (EBL by Elionix, ELS-HS50). The gold and chromium layers are then deposited with an electron-beam evaporator (AJA, Orion-8E), with a thickness of 60 nm and 30 nm, respectively. For details on our fabrication process, please see our previous work \cite{chen2020efficient}. The waveguide areas are then periodically poled by applying high-voltage pulses onto the electrode pads. Once the poling has been completed, the electrodes and pads are removed in a metal etcher solution. \\
 
Next, the waveguide pattern is defined using another EBL at 100 keV (Elionix ELS-G100), after the 850 nm hydrogen silsesquioxane (HSQ) resist is spun on the thin film; see the detailed writing process in \cite{chen2018modal}. After development, it is moved to etching by inductively coupled plasma etch-chlorine (ICP-CL, Oxford PlasmaPro System 100 Cobra). The etching depth is set to 350 nm, resulting in a 250 nm remaining slab layer. Next, the RCA-1 bath and buffered-oxide etch (BOE, 6:1) are applied to clean up re-deposition and remove the remaining resist. The sidewall is then checked using a scanning electron microscope (SEM, FEI-NOVA NANO). Lastly, a thin layer of silicon dioxide is applied using plasma-enhanced chemical vapor deposition (PECVD, Oxford Plasmapro NPG80), for protection from damage. \\
 
The result is a periodic poled lithium niobate (PPLN) nano-waveguide with 1.4~$\mu$m top width, 350 nm etching depth, and $64^{\circ}$ side-wall angle, and 1.71 $\mu$m poling period, as shown in Fig.~\ref{fig:PPLN Blue Light pole check}. 
Along the waveguide sidewall there are contrast-shaded sectors (true-color under SEM), where the contrast shows the poled sectors (each lighter sector 0.655 $\mu$m wide, each darker sector 1.08 $\mu$m wide). As such, the poling duty cycle (ratio of poled to unpoled) is 35:65. The measured poling period is 1.706 $\mu$m, which is very close to the designed period of 1.71 $\mu$m.

\section{Experiment}

\begin{figure}[ht]
\centering
\includegraphics[width=\linewidth]{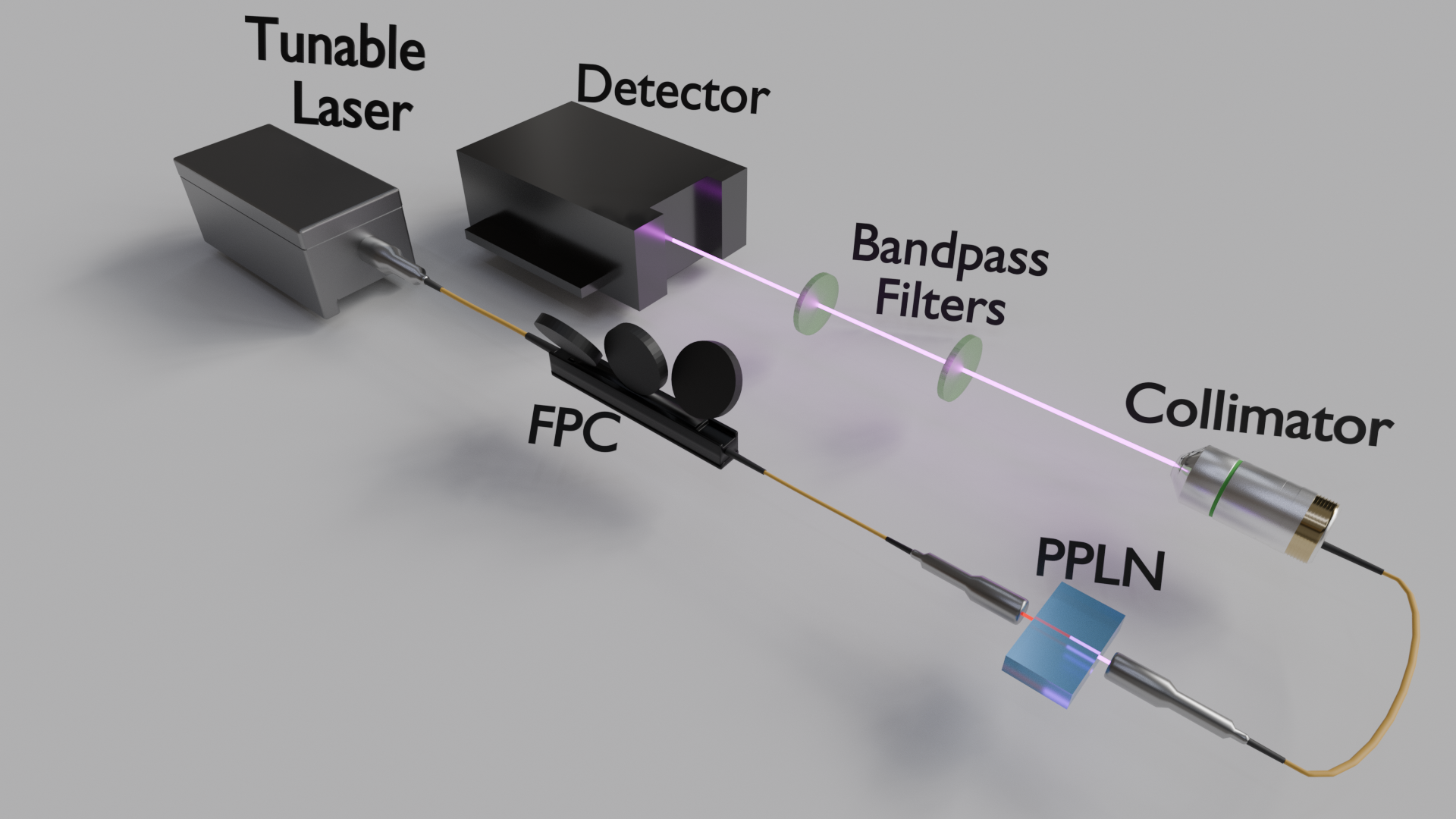}
\caption{Experimental setup, with a laser, an FPC, the PPLN chip coupled to lensed fibers, an aspheric lens as the collimator to free space, two band-pass filters, and a UV-light detector.}
\label{fig:Blue Light Set Up-Paper}
\end{figure} 

The experimental setup is shown in Fig.~\ref{fig:Blue Light Set Up-Paper}. It begins with a wavelength-tunable laser (New Focus, 765-781 nm), whose output is sent through a fiber polarization controller (FPC) and coupled onto the PPLN chip through a lensed fiber (OZ Optics TSMJ-3A-780), with a measured 7.67 dB loss. The chip is housed on a temperature controlled stage stabilized at $35^{\circ}$C. At the chip output, the light is collected by another lensed fiber (loss: 7.67 dB) and launched into free space through a collimator with a 400 nm aspheric lens (Thorlabs, N414TM). Then, two band-pass filters (Thorlabs, FBH400-40) are inserted to reject the pump, after which the generated UV-A is measured using a power meter (Thorlabs, S130VC). 

\begin{figure}[ht]
\centering
\includegraphics[width=\linewidth]{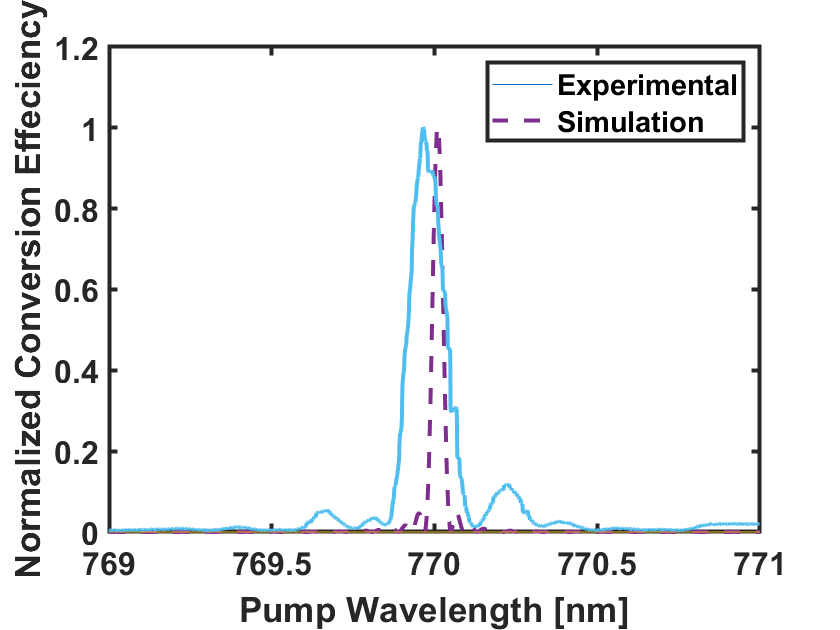}
\caption{Normalized phase matching curves for UV-A generated light in PPLN waveguide.Experimental (solid) and simulated (dashed) are normalized to 1 for comparison.}
\label{fig:Blue light phase match}
\end{figure}

To characterize the waveguide's quasi-phase matching properties, the wavelength of the laser is coarse scanned over its entire spectral range (765-781 nm) while the UV-A power is measured. The peak generated power is about 770 nW, found at around 770.08 nm. Then, the measurement is repeated with the laser fine scanning from 769 to 771 nm, while its output power is kept at 12 mW. The recorded phase matching curve is shown in Fig.~\ref{fig:Blue light phase match}, with the normalized conversion efficiency plotted against the wavelength. As seen, the measured phase matching bandwidth has a full width at half maximum (FWHM) of 0.3 nm. In contrast, the simulated FWHM is only 0.1 nm, which indicates the different group velocity dispersion from simulation than experiment, due to fabrication imperfections. \\ 

Next, a 400-nm laser beam (QPhotonics Laser Diode ) is transmitted through the optics setup, to measure the on/off chip losses, and the insertion losses of the collimator and filters. The total insertion loss from fiber to fiber is measured to be 21.65 dB, which is rather high because the lensed fibers used for the chip coupling are for 780 nm. In view of the symmetric coupling setup, the on-chip and off-chip losses are assumed to be the same, which is 10.83 dB per facet. Afterwards, the total loss of free space components is measured to be 0.3 dB, by transmitting the same laser beam through them directly. \\
\begin{figure}[ht]
\centering
\includegraphics[width=\linewidth]{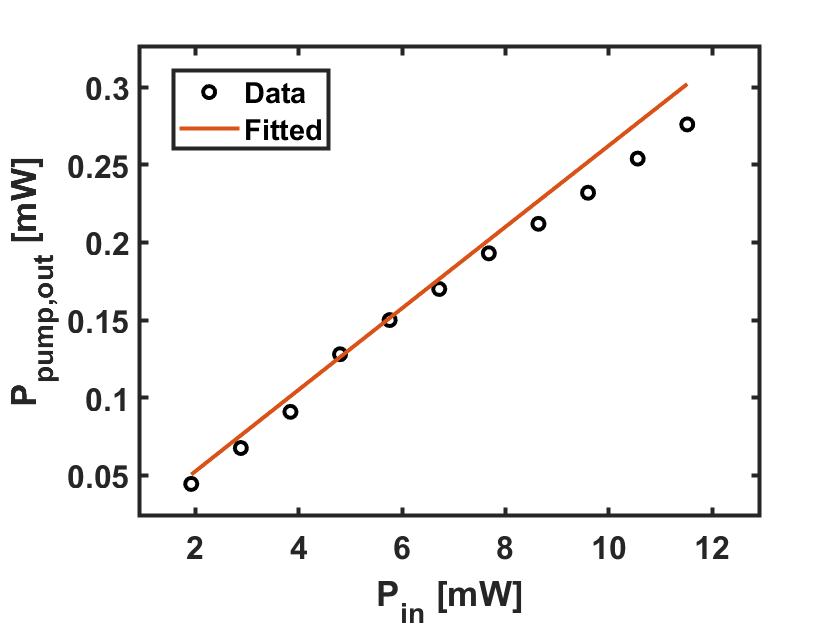}
\caption{The measured pump power collected by the lensed fiber at the chip output as its input power is swept (circles), and its quadratic fit (solid).}
\label{fig:red light power sweep}
\end{figure}
\begin{figure}[ht]
\centering
\includegraphics[width=\linewidth]{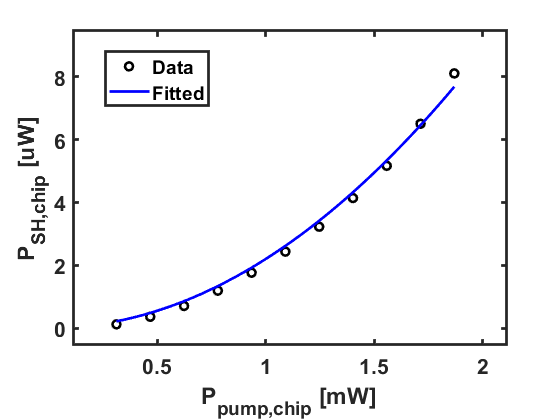}
\caption{The measured second-harmonic power collected by the lensed fiber at the chip output as its input power is swept (circles), along with its quadratic fit (solid).}
\label{fig:Blue light power sweep}
\end{figure}

To measure the conversion efficiency, the input pump power is first set at 2 mW and its wavelength fine scanned from 769 nm to 771 nm to park at the maximum conversion wavelength. Then, the output powers of the pump and SH light are measured over two consecutive runs as the input pump power is swept from 2 mW to 12 mW. Figure~\ref{fig:red light power sweep} plots the output pump power $P_\mathrm{pump, out}$, as a function of the input power $P_\mathrm{in}$, each measured at fibers before and after the chip. As shown, the dependence is mostly linear with only a slight deviation, which indicates weak SHG. Fitting the data with
\begin{equation} 
P_\mathrm{pump, out}=\beta(P_\mathrm{in}-\alpha P_\mathrm{in}^2)
\end{equation}
finds $\beta=(2.63\pm 0.18)\times 10^{-2}$ and $\alpha=(7.7\pm 6.9)\times 10^{-3}$/mW, where the uncertainties correspond to a confidence level of 95\%. Here, $\beta$ measures the insertion loss, which is converted to a 15.8 dB loss, consistent with our separately measured loss of 15.3 dB. $\alpha$ is proportional to the conversion efficiency. Due to the weak SHG, however, the uncertainty from the curve fitting is too high. \\

To reliably extract the conversion efficiency, we instead use the $\beta$ value from Fig.~\ref{fig:red light power sweep} to fit the generated SH power as 
\begin{equation}
    P_\mathrm{SH, chip}=\gamma P_\mathrm{pump, chip}^2
\end{equation}
Here, $P_\mathrm{SH, chip}$ is the on-chip SH power, calculated by multiplying the measured SH power by $10^{1.098}$, to account for the off-chip loss and transmission loss in free space. $P_\mathrm{pump, chip}$ is the on-chip pump power calculated as $\sqrt{\beta}P_\mathrm{in}$, to account for the on-chip loss. The results are shown in Fig.~\ref{fig:Blue light power sweep}, where a clear quadratic behavior is seen, as opposed to previously reported \cite{hwang2023tunable}. The fitting gives $\gamma=(2.1\pm 0.1)\times 10^{-3}$/mW, with which the normalized efficiency is computed as $\gamma/L^2$ to give $\eta=1797 \% W^{-1} cm^{-2}$. Note that this efficiency value hinges on the accuracy of the 400-nm light loss measurement, which we repeatedly performed to verify our results. \\

In comparison, our simulated efficiency for the 35:65 poling duty cycle is found to be 2905 $\% W^{-1} cm^{-2}$, which is 1.6 times higher than the experimental results. This discrepancy is partially because our simulation has not taken into account the waveguide propagation loss. It also signifies the imperfections in our fabrication, including non-uniform poling patterns and that the side wall angle deviates from the design.

\section{Conclusion}
We have presented efficient UV-A light generation on chip, with a record high conversion efficiency of 1797 $\% W^{-1} cm^{-2}$, corresponding to a 9.1 times improvement over the state-of-the-art; as listed in the table below. Our improvement is attributed to tighter mode confinement, thanks to a smaller top width and etching depth, and better poling quality. \\ % "previously" instead of "theirs:"

\begin{table}[hbt!]
\centering
\begin{tabular}{ccc}
\hline
Refs \; &  Generation Wavelength \;  &  Efficiency \\
\hline
 \cite{park2022high}&  $456.5$ nm & $ 33000 \% W^{-1} cm^{-2}$ \\
 \cite{sayem2021efficient}&  $435.5$ nm& $\sim 1040 \% W^{-1} cm^{-2}$ \\
 \cite{hwang2023tunable}&  $355$ nm& $\sim 197 \% W^{-1} cm^{-2}$ \\
 This work & $390$ nm& $1797 \% W^{-1} cm^{-2}$ \\
\hline
\end{tabular}
\caption{Comparison of near UV SHG by various sources.}
  \label{tab:results compare}
\end{table}
\smallskip

For future improvement, better poling quality can be achieved by perfecting the fabrication of the electrodes and the applied voltage duration. For even shorter wavelength generation, the periodic poling may need to be reduced. Our results and further developments would open doors to powerful applications in atomic clocks, quantum memory, quantum sensing, and so on, where photonic integrated circuits could provide much practical values.

\section{Acknowledgment} 
The fabrication was performed in part at the Advanced Science Research Center Nanofabrication Facility of the Graduate Center at the City University of New York. This work is supported in part by the ACC-New Jersey under Contract No. W15QKN-18-D-0040.

\section{Disclosures}The authors declare no conflicts of interest.

\section{Author Note}
This experiment was completed in October 2024. After completing the draft we became aware of Ref.~\cite{Franken2025Milliwatt-level}, a pre-print that also reports UV generation on thin film lithium niobate. 
%arXiv:2503.16785  https://arxiv.org/pdf/2503.16785. 

\smallskip

\section{References}

% Bibliography
\bibliography{BlueLightGeneration}

\begin{thebibliography}{99}

\bibitem{hou2024high}
Songyan Hou, Hao Hu, Zhihong Liu, Weichuan Xing, Jincheng Zhang, Yue Hao, "High-Speed Electro-Optic Modulators Based on Thin-Film Lithium Niobate", \textit{Nanomaterials}, vol. 14, no. 10, p. 867, 2024.

\bibitem{feng2024integrated}
Hanke Feng, Tong Ge, Xiaoqing Guo, Benshan Wang, Yiwen Zhang, Zhaoxi Chen, Sha Zhu, Ke Zhang, Wenzhao Sun, Chaoran Huang, \textit{et al.}, "Integrated lithium niobate microwave photonic processing engine", \textit{Nature}, vol. 627, no. 8002, pp. 80--87, 2024.

\bibitem{shi2024efficient}
Xiaodong Shi, Sakthi Sanjeev Mohanraj, Veerendra Dhyani, Angela Anna Baiju, Sihao Wang, Jiapeng Sun, Lin Zhou, Anna Paterova, Victor Leong, Di Zhu, "Efficient photon-pair generation in layer-poled lithium niobate nanophotonic waveguides", \textit{Light: Science \& Applications}, vol. 13, no. 1, p. 282, 2024.

\bibitem{tang2024broadband}
Chao Tang, Mingming Nie, Jia-yang Chen, Zhaohui Ma, Zhan Li, Yijun Xie, Yong Meng Sua, Shu-Wei Huang, Yu-Ping Huang, "Broadband frequency comb generation through cascaded quadratic nonlinearity in thin-film lithium niobate microresonators", \textit{Optics Letters}, vol. 49, no. 9, pp. 2449--2452, 2024.

\bibitem{hwang2023tunable}
Emily Hwang, Nathan Harper, Ryoto Sekine, Luis Ledezma, Alireza Marandi, Scott Cushing, "Tunable and efficient ultraviolet generation with periodically poled lithium niobate", \textit{Optics Letters}, vol. 48, no. 15, pp. 3917--3920, 2023.

\bibitem{chen2018modal}
Jia-Yang Chen, Yong Meng Sua, Heng Fan, Yu-Ping Huang, "Modal phase matched lithium niobate nanocircuits for integrated nonlinear photonics", \textit{OSA Continuum}, vol. 1, no. 1, pp. 229--242, 2018.

\bibitem{chen2020efficient}
Jia-Yang Chen, Chao Tang, Zhao-Hui Ma, Zhan Li, Yong Meng Sua, Yu-Ping Huang, "Efficient and highly tunable second-harmonic generation in Z-cut periodically poled lithium niobate nanowaveguides", \textit{Optics Letters}, vol. 45, no. 13, pp. 3789--3792, 2020.

\bibitem{luo2019semi}
Rui Luo, Yang He, Hanxiao Liang, Mingxiao Li, Qiang Lin, "Semi-nonlinear nanophotonic waveguides for highly efficient second-harmonic generation", \textit{Laser \& Photonics Reviews}, vol. 13, no. 3, p. 1800288, 2019.

\bibitem{wang2021single}
Pengfei Wang, Chun-Yang Luan, Mu Qiao, Mark Um, Junhua Zhang, Ye Wang, Xiao Yuan, Mile Gu, Jingning Zhang, Kihwan Kim, "Single ion qubit with estimated coherence time exceeding one hour", \textit{Nature communications}, vol. 12, no. 1, p. 233, 2021.

\bibitem{liu2019beyond}
Xianwen Liu, Alexander W Bruch, Juanjuan Lu, Zheng Gong, Joshua B Surya, Liang Zhang, Junxi Wang, Jianchang Yan, Hong X Tang, "Beyond 100 THz-spanning ultraviolet frequency combs in a non-centrosymmetric crystalline waveguide", \textit{Nature communications}, vol. 10, no. 1, p. 2971, 2019.

\bibitem{park2022high}
Taewon Park, Hubert S Stokowski, Vahid Ansari, Timothy P McKenna, Alexander Y Hwang, MM Fejer, Amir H Safavi-Naeini, "High-efficiency second harmonic generation of blue light on thin-film lithium niobate", \textit{Optics Letters}, vol. 47, no. 11, pp. 2706--2709, 2022.

\bibitem{sayem2021efficient}
Ayed Al Sayem, Yubo Wang, Juanjuan Lu, Xianwen Liu, Alexander W Bruch, Hong X Tang, "Efficient and tunable blue light generation using lithium niobate nonlinear photonics", \textit{Applied Physics Letters}, vol. 119, no. 23, 2021.

\bibitem{yoon2017coherent}
Dong Yoon Oh, Ki Youl Yang, Connor Fredrick, Gabriel Ycas, Scott A Diddams, Kerry J Vahala, "Coherent ultra-violet to near-infrared generation in silica ridge waveguides", \textit{Nature communications}, vol. 8, no. 1, p. 13922, 2017.

\bibitem{xue2023chip}
Miao Xue, Xiongshuo Yan, Jiangwei Wu, Rui Ge, Tingge Yuan, Yuping Chen, Xianfeng Chen, "On-chip ultraviolet second-harmonic generation in lithium-tantalate thin film microdisk", \textit{Chinese Optics Letters}, vol. 21, no. 6, p. 061902, 2023.

\bibitem{lu2019periodically}
Juanjuan Lu, Joshua B Surya, Xianwen Liu, Alexander W Bruch, Zheng Gong, Yuntao Xu, Hong X Tang, "Periodically poled thin-film lithium niobate microring resonators with a second-harmonic generation efficiency of 250,000\%/W", \textit{Optica}, vol. 6, no. 12, pp. 1455--1460, 2019.

\bibitem{jin2021efficient}
Mingwei Jin, Jiayang Chen, Yongmeng Sua, Prajnesh Kumar, Yuping Huang, "Efficient electro-optical modulation on thin-film lithium niobate", \textit{Optics letters}, vol. 46, no. 8, pp. 1884--1887, 2021.

\bibitem{ma2020ultrabright}
Zhaohui Ma, Jia-Yang Chen, Zhan Li, Chao Tang, Yong Meng Sua, Heng Fan, Yu-Ping Huang, "Ultrabright quantum photon sources on chip", \textit{Physical Review Letters}, vol. 125, no. 26, p. 263602, 2020.

\bibitem{fan2021photon}
Heng Fan, Zhaohui Ma, Jia-Yang Chen, Zhan Li, Chao Tang, Yong Meng Sua, Yuping Huang, "Photon conversion in thin-film lithium niobate nanowaveguides: a noise analysis", \textit{JOSA B}, vol. 38, no. 7, pp. 2172--2179, 2021.

\bibitem{li2020lithium}
Mingxiao Li, Jingwei Ling, Yang He, Usman A Javid, Shixin Xue, Qiang Lin, "Lithium niobate photonic-crystal electro-optic modulator", \textit{Nature Communications}, vol. 11, no. 1, p. 4123, 2020.

\bibitem{wang2018nanophotonic}
Cheng Wang, Mian Zhang, Brian Stern, Michal Lipson, Marko Lon{\v{c}}ar, "Nanophotonic lithium niobate electro-optic modulators", \textit{Optics express}, vol. 26, no. 2, pp. 1547--1555, 2018.

\bibitem{chen2019efficient}
Jia-yang Chen, Yong Meng Sua, Zhao-hui Ma, Chao Tang, Zhan Li, Yu-ping Huang, "Efficient parametric frequency conversion in lithium niobate nanophotonic chips", \textit{Osa Continuum}, vol. 2, no. 10, pp. 2914--2924, 2019.


\bibitem{wu2024visible}
Tsung-Han Wu, Luis Ledezma, Connor Fredrick, Pooja Sekhar, Ryoto Sekine, Qiushi Guo, Ryan M Briggs, Alireza Marandi, Scott A Diddams, "Visible-to-ultraviolet frequency comb generation in lithium niobate nanophotonic waveguides", \textit{Nature Photonics}, vol. 18, no. 3, pp. 218--223, 2024.

\bibitem{edamatsu2004generation}
Keiichi Edamatsu, Goro Oohata, Ryosuke Shimizu, Tadashi Itoh, "Generation of ultraviolet entangled photons in a semiconductor", \textit{Nature}, vol. 431, no. 7005, pp. 167--170, 2004.

\bibitem{king2022optical}
Steven A King, Lukas J Spie{\ss}, Peter Micke, Alexander Wilzewski, Tobias Leopold, Erik Benkler, Richard Lange, Nils Huntemann, Andrey Surzhykov, Vladimir A Yerokhin, \textit{et al.}, "An optical atomic clock based on a highly charged ion", \textit{Nature}, vol. 611, no. 7934, pp. 43--47, 2022.

\bibitem{Franken2025Milliwatt-level}
C. A. A. Franken, S. S. Ghosh, C. C. Rodrigues, J. Yang, C. J. Xin, S. Lu, D. Witt, G. Joe, G. S. Wiederhecker, K. -J. Boller, M. Lončar,
\textit{et al.}, "Milliwatt-level UV generation using sideall poled lithium niobate", \textit{arxiv}, 2025.

\end{thebibliography}
 \bibliographyfullrefs{BlueLightGeneration}
{}

\end{document}